\documentstyle[preprint,eqsecnum,aps]{revtex}

\begin{document}
\draft
\title{Spectroscopic Factors in $^{40}$Ca and $^{208}$Pb
from $(e,e'p)$: Fully Relativistic Analysis}

\author{J.M. Ud\'\i as, P. Sarriguren, E. Moya de Guerra, E. Garrido
and J.A. Caballero}
\address{
Instituto de Estructura de la Materia, \\
Consejo Superior de Investigaciones Cient\'\i ficas, \\
Serrano 119, E-28006 Madrid, Spain\\
}
\maketitle
\begin{abstract}
We present results for spectroscopic factors of the  outermost
shells in $^{40}$Ca and $^{208}$Pb, which have been derived from
the comparison between the available quasielastic ($e,e'p$) data
from NIKHEF-K and the corresponding calculated cross-sections
obtained within a fully relativistic formalism.
We include exactly the effect of Coulomb distortion on the electron
wave functions and discuss its role in the extraction of the
spectroscopic factors from experiment. Without any adjustable
parameter, we find spectroscopic factors of about 70\%, consistent
with theoretical predictions. We compare our results with previous
relativistic and nonrelativistic analyses of ($e,e'p$) data. In
addition to Coulomb distortion effects we discuss different choices
of the nucleon current operator and also analyze the effects due to
the relativistic treatment of the outgoing-distorted and bound
nucleon wave functions.

\end{abstract}
\pacs{25.30.-c, 24.10.Jv, 21.10.Jx}

\section{Introduction}
\label{sec:intro}

Spectroscopic factors and occupation probabilities are basic
elements for our understanding of the nuclear structure, measuring
the accuracy of the shell model description. The fundamental
concept on which the shell model is based, the mean field
approximation, is considered as the leading contribution in an
expansion of multiparticle correlations within the general framework
of the nuclear many-body theory. These correlations allow the
nucleons to occupy partially different orbits above and below the
Fermi level. Therefore, the deviation from full occupancy of the
orbits below the Fermi level is a measure of the correlations
neglected in the mean field approach, or in other words, a check of
the validity of the mean field description.

The spectroscopic factor $S_{\alpha}$ is defined as the probability
to reach a final single particle (hole) state $\alpha$ when a
nucleon is added to (removed from) the target nucleus. The
occupation number $N_{\alpha}$ is the number of nucleons in the
quantum state $\alpha$ in the target nucleus, relative to the
$2j+1$ limit.

Theoretical spectroscopic factors and occupation numbers have been
derived for some doubly magic nuclei, taking into account both
short- and long-range correlations. In particular, occupation
numbers for the $3s_{1/2}$ state in $^{208}$Pb have been obtained
from different approaches. Mahaux and Sartor \cite{Mah91} derived
an occupation number $N_{3s}=0.83$  and a spectroscopic factor
$S_{3s}=0.69$ by using a dispersion relation approach, which allows
to extrapolate the empirical mean field from positive energies
(optical potential) to negative energies (shell model potential).
Slightly smaller spectroscopic factors have been obtained by Ma and
Wambach \cite{Ma91} using a quasiparticle hamiltonian which includes
correlations in a phenomenological way. Pandharipande and
collaborators \cite{Pan84} obtained an average value $0.71\pm 0.1$
for the occupation probabilities of single-particle states just
below the Fermi level, using a variational calculation of nuclear
matter to which a random phase approximation correction is added.

{}From the experimental point of view, relative spectroscopic factors
have been historically determined by single-nucleon transfer
reactions. In a search for absolute empirical values of spectroscopic
factors, additional experimental information has been recently
collected from two new sources. One is based on a sum rule analysis
\cite{sumrule} of both transfer data and charge density differences
of isotones extracted from $(e,e')$ reactions. This method allows
to go from relative spectroscopic factors to an absolute occupation
number (see Ref. \cite{Sic91} and Refs. therein). A typical value
for the occupation number of the $3s_{1/2}$ shell in $^{208}$Pb
obtained from this method is $N_{3s}=0.78\pm 0.12$ \cite{Sic91}.

The other method is the quasielastic $(e,e'p)$ reaction, which
constitutes indeed a very well suited tool to extract experimental
information on absolute spectroscopic factors. Modern facilities
allow to study this reaction in detail and high precision
measurements of cross sections are available
\cite{Quint,Kra89,deW90}. The small strength of the electromagnetic
interaction allows one to study the transition in Born approximation
(BA) by the exchange of a single virtual photon. Choosing properly
the momentum and energy transferred by the virtual photon in order
to satisfy the quasielastic condition $(\omega \simeq q^2/2M)$,
the process can be treated with confidence \cite{Fru85} in Impulse
Approximation (IA), i.e., assuming that the exchanged photon is
absorbed by a single nucleon which is the one detected. Under
these circumstances, it is possible to extract information on the
energy and momentum distribution of the bound nucleon.

Although the advantages of the $(e,e'p)$ process over other possible
reactions to study spectroscopic factors are widely recognized, the
extraction of these factors from experiment is still not free of
ambiguities. A reliable determination of spectroscopic factors
requires an accurate knowledge of the mechanism of the reaction and,
in this context, the exact treatment of the  Coulomb distortion of
the electrons is important, especially in heavy nuclei such as lead.

Different methods have been proposed to handle the distortion. The
early approaches were based on the eikonal approximation \cite{GP}
within a nonrelativistic scenario for the nuclear part. This method
was applied to the analysis of $(e,e'p)$ data taken at NIKHEF-K
\cite{Quint,Kra89}, producing surprisingly low spectroscopic factors
($0.49\pm 0.05$ for the $3s_{1/2}$ shell in $^{208}$Pb), incompatible
with results obtained from other reactions as well as with
theoretical predictions. This fact, together with the large
difference between the results from first and second order eikonal
calculations, raised the question of whether this approximation was
adequate to treat the Coulomb distortion or a more involved analysis
was still needed.

The first realistic calculations with a more exact treatment of the
Coulomb distortion in ($e,e'p$) were made in 1990 by  McDermott
\cite{McD90}. The incorporation of the  distortion effects was
achieved by means of a full partial wave analysis of the electron
waves in the Coulomb potential of the target, as well as of the
outgoing proton waves distorted by the optical potential. The
complexity of the numerical calculations required a full relativistic
framework, not only for the electron vertex where the energies
involved in the experiments (hundreds of MeV) make it mandatory, but
also for the nuclear vertex. Nevertheless, some approximations were
still made (helicity conserved approximation, HCA) to simplify the
treatment of the electron Coulomb distorted waves. The spectroscopic
factor for the $3s_{1/2}$ shell in $^{208}$Pb amounts in this
approximation to $0.65$ \cite{McD90}, manifestly above the value
obtained within the eikonal approximation.

Subsequently, a new relativistic calculation was reported by Jin
{\em et al.} \cite{Jin92}, where the treatment of the Coulomb
distortion of the electrons was exact and the above mentioned
restricted approximation (HCA) was not made. The result of this
calculation \cite{Jin92} for  the spectroscopic factor in the same
shell is $0.71$, which is larger than the result of Ref.
\cite{McD90}. The authors of Ref. \cite{Jin92} reported also some
discrepancies with the results of Ref. \cite{McD90} in the limit
of plane waves for the electron (no Coulomb distortion) and stressed
the need for further investigation.

This paper is an attempt to clarify this situation by an independent
analysis including the development of a new code \cite{Udi93}, which
treats the Coulomb distortion of the electron in an exact way and
uses a relativistic formalism for both the leptonic and the nuclear
vertices. Our calculations are along similar lines to those of Ref.
\cite{Jin92}, though in some instances we use a different nucleon
current operator and different bound nucleon wave functions.
Differences and similarities with previous calculations in Refs.
\cite{McD90,Jin92} are discussed in detail in the next sections,
which are organized as follows: In Sec. II we summarize briefly
the formalism involved in our calculations. In Sec. III we discuss
our results for the $3s_{1/2}$ and $2d_{3/2}$ shells in $^{208}$Pb
and for the $2s_{1/2}$ and $1d_{3/2}$ shells in $^{40}$Ca. We
discuss not only the effect of Coulomb distortion but also the
effect of the relativistic optical potential and of nucleon current
operator. Sec. IV summarizes the main conclusions.

\section{Description of calculations}
\label{sec:form}

In this section we summarize the formalism used to describe the
coincidence $(e,e'p)$ reaction under the conditions defining the
IA discussed in the Introduction. Fig. 1
represents graphically the $(e,e'p)$ process. In this figure
$k_i^\mu$ ($k_f^\mu$) is the four-momentum of the incoming (outgoing)
electron and $q^\mu$ is the four-momentum of the exchanged photon.
The ejected proton four-momentum is denoted by $P_F^\mu$. We use the
notation and conventions of Ref. \cite{BD64} as well as $\hbar=c=1$.
As represented in Fig. 1 the electromagnetic transition is treated
in Born approximation, although we take into account the effect of
the nuclear Coulomb potential to all orders by using electron
distorted wave functions.

The differential cross-section for this process is then written as
\cite{Udi93}:
\begin{eqnarray}
\frac{d\sigma}{d\epsilon_f d\Omega_f d E_F d \Omega_F}&=&
\frac{\delta(\epsilon_i+E_A-\epsilon_f-E_F-E_{A-1})} {(2\pi)^5}
\nonumber
\\
\\  \nonumber
&&\times 4 \alpha^2 \epsilon^{2}_{f}  E_F |\vec{P}_F|\
\overline{\sum} |W_{if}|^2 \; ,
\label{eqb7}
\end{eqnarray}
where $\overline{\Sigma}$ indicates sum (average) over final
(initial) polarizations and
\begin{equation}
W_{if}=\int\!\!\!  d\vec{x} \int\!\!\!
d\vec{y} \int\!\!\!  \frac{d\vec{q}}{(2\pi)^{2}}
j_\mu^e(\vec{x})e^{-i\vec{q}(\vec{x}-\vec{y})}\frac{(-1)}{q_\mu^2}
J_N^\mu(\vec{y})\; .
\label{eqb8}
\end{equation}
In this expression $j_\mu^e$ and $J_N^\mu$ stand for the electron
and nuclear currents, respectively. The electron current is given
by the well known point-like Dirac particle expression:
\begin{equation}
j^\mu_e(\vec{r})=
\bar{\psi}^e_f (\vec{r}) \gamma^\mu \psi^e_i(\vec{r}) \; ,
\label{eqb9}
\end{equation}
where  $\psi^e_i, \psi^e_f$ stand for initial and final electron
wave functions. In IA and within an independent
particle model picture, the nuclear current can be written in terms
of the nucleon current operator $ \hat{J}^\mu _N $
\begin{equation}
J^\mu_N(\vec{r})=
\bar{\psi}^N_F (\vec{r}) \hat{J}^\mu _N \psi^N_B(\vec{r}) \; ,
\end{equation}
with $\psi^N_B, \psi^N_F$ the wave functions for the initial bound
nucleon and final nucleon, respectively, and $\hat{J}^\mu _N$ a
nucleon current operator to be specified later.

The initial and final electron wave functions, solutions of the
Dirac equation with the Coulomb potential, have the form
\begin{eqnarray}
\psi^e_i(\vec{r})=4 \pi \sqrt{\frac{\epsilon_i+m}{2\epsilon_i}}
&{\displaystyle \sum_{\kappa,\mu,m}}&
 e^{i\delta_{\kappa}}
i^{l} < l \ m \ \frac{1}{2} \ \sigma_i | j \ \mu >\nonumber \\
&& \times Y_{lm}^{*}(\hat{k}_i)\psi_{\kappa}^{\mu}(\vec{r}) \; ,
\label{expan}
\end{eqnarray}
where the functions
\begin{equation}
\psi^{\mu}_{\kappa}(\vec{r})=\left(\begin{array}
{@{\hspace{0pt}}c@{\hspace{0pt}}}
g_{\kappa}(r)
\phi^\mu_\kappa(\hat{r}) \\
 i f_{\kappa}(r) \phi^\mu_{-\kappa}(\hat{r})\end{array}\right)
\label{eqa8}
\end{equation}
are eigenstates of total angular momentum with quantum numbers
$\kappa \mu$ ($j=|\kappa|-1/2$; $l=\kappa$ if $\kappa>0$ and
$l=-\kappa-1$ if $\kappa<0$).  The functions $f_{\kappa},g_{\kappa}$
satisfy the usual radial equations \cite{UbeRose} and
$\phi^{\mu}_{\kappa}(\hat{r})$ is given by
\begin{equation}
\phi^{\mu}_{\kappa}(\hat{r}) =[Y_{l} \otimes
\vec{\sigma}]^{\mu}_{j}\equiv \sum_{m,\sigma}
< l \ m \ \frac{1}{2} \ \sigma | \ j \ \mu >  Y_{lm}(\hat{r})
\chi_{\sigma}^{\frac{1}{2}}\; .
\end{equation}
For the outgoing electron wave function $\psi^e_f$, the phase shifts
$\delta_\kappa$ have to be included with a minus sign. The functions
$f_{\kappa},g_{\kappa}$ and the phase shifts are obtained by
numerical integration of the radial equations using the Milne
procedure, as described in the work of Yennie {\em et al.}
\cite{Yen}, including up to third derivatives. The Coulomb potentials
for $^{208}$Pb and $^{40}$Ca are derived from the experimental charge
distributions given in Ref. \cite{deV}.

The bound state wave functions for the proton $\psi^N_B$ are spinors
with well defined angular momentum quantum numbers $\kappa_B \mu_B$,
and have a structure similar to that in Eq. (\ref{eqa8}).  They have
been computed within the framework of the relativistic
independent particle shell model. The mean field in the Dirac
equation is determined through a Hartree procedure from a
phenomenological relativistic lagrangian with scalar and vector S-V
terms. We use the parameters of Ref. \cite{HS}, which are fitted to
reproduce nuclear matter properties and the charge radius in
$^{40}$Ca, and the TIMORA code \cite{HSbook}.

The wave function of the detected proton $\psi^N_F$ is a scattering
solution of a Dirac-like equation, which includes S-V global optical
potentials, obtained by fitting elastic proton scattering data
\cite{Hama}. This wave function has basically the same structure of
Eq. (\ref{expan}) except that, since the potential is in
this case complex, the phase shifts and radial functions are also
complex. In addition, since the wave function corresponds to an
outgoing proton we have to use in Eq. (\ref{eqa8}) the complex
conjugates $f_\kappa ^\star, g_\kappa ^\star $ and in Eq.
(\ref{expan}) the complex conjugate $\delta^\star$ with a negative
sign.

For the current operator we consider the two choices $cc1$ and $cc2$
introduced by de Forest \cite{deF83} in momentum space
\begin{equation}
\hat{J}_{N}^\mu (cc1)=(F_1+\bar{\kappa}F_2)\gamma^\mu-\frac{\bar
{\kappa}F_2}{2M}(P_F+\bar{P}_I)^\mu \; ,
\label{cc1}
\end{equation}
\begin{equation}
\hat{J}_{N}^\mu (cc2)=F_1\gamma^\mu+i \frac{\bar{\kappa}F_2}{2M}
\sigma^{\mu\nu}q_\nu \; ,
\label{cc2}
\end{equation}
where $F_1$ and  $F_2$ are the nucleon form factors related in the
usual way \cite{BD64} to the electric and magnetic Sachs form
factors of the dipole form.  $\bar{P}_I$ in Eq. (\ref{cc1}) is the
four-momentum of the initial nucleon assuming on-shell kinematics
\cite{deF83}.

As it is well known \cite{BD64}, Eqs. (\ref{cc1}) and (\ref{cc2})
are equivalent when the initial and final nucleons are on-shell.
For off-shell nucleons, as is our case, both expressions lead to
different results and do not satisfy current conservation.
In configuration space all three-momenta in Eqs. (\ref{cc1}) and
(\ref{cc2}) have to be considered as operators. For the $cc2$
choice, integration by parts allows one to replace the gradient
operators acting on $\psi^N_B, \psi^N_F$ by $\vec{q}$, the variable
of integration in Eq. (\ref{eqb8}). In the case of the $cc1$ choice,
it is not possible to get rid of the gradient operator acting on at
least one of the nucleon wave functions. Given the fact that for
off-shell nucleons none of the expressions ($cc1,cc2$) is fully
satisfactory and both expressions fail to verify current
conservation, when using the choice $cc1$ we use the simplifying
assumption of replacing $\hat{\vec{P}}_F + \hat{\vec{P}}_I$  by
the asymptotic $\vec{P}_F + \vec{P}_m$ values, with $\vec{P}_m$
the missing momentum. This assumption simplifies enormously the
calculations and is consistent with the prescription of de Forest
in Ref. \cite{deF83} for the half-off-shell electron-proton
cross-section $\sigma (cc1)$, which is commonly used. In the same
spirit $F_1$ and $F_2$ are taken as the standard nucleon form
factors at the asymptotic $q$-values, an approximation usually
made when Coulomb distorted electron waves are employed
\cite{McD90,Jin92}. Under this approximation the computation time
is highly reduced.

The operator $cc2$ was used in previous fully  relativistic
calculations \cite{McD90,Jin92}. In this work we use both $cc1$
and $cc2$ operators.
Comparisons between half-off-shell electron-proton  cross-sections
obtained with both operators ($cc1$ and $cc2$) have previously been
made using plane waves for initial and final nucleons
\cite{deF83,Cab93}. In Refs. \cite{deF83} and \cite{Cab93} small
differences  (a few percent) were found at the kinematical
conditions usually attained in the experiments. In the next section
we present similar comparisons when realistic spinors are used to
obtain the cross-section.

Different notations have been introduced in the literature by
different authors to distinguish various degrees of approximations
to the reaction mechanism. To avoid confusion we shall specify
separately the approximation taken at each vertex, i.e., at the
electronic and at the nuclear vertices. To distinguish the case
when we use plane waves from the case when we use distorted waves
in the electron current (Eq. (\ref{eqb9})), we use the notations
PWBA and DWBA, respectively. On the other hand, to distinguish the
cases when we use plane waves and distorted waves for the ejected
proton, we use the notation PWIA and DWIA, respectively. Unless
otherwise specified, the results presented are obtained with proton
distorted waves (DWIA). In PWBA the electron current is given by
\begin{equation}
j^\mu_{e,free}(\vec{r})=\frac{m}{\sqrt{\epsilon_i\epsilon_f}}
e^{i(\vec{k}_i-\vec{k}_f)\vec{r}} \bar{u}(\vec{k}_f,\sigma_f)
\gamma^\mu u(\vec{k}_i,\sigma_i) \; .
\end{equation}
In the same way, when the optical potential is not included in the
calculations (PWIA), the wave function of the ejected proton
$\psi^N_F$  becomes a plane wave in the corresponding equation for
the nuclear current.

A few remarks concerning the numerical calculations in DWBA are in
order. The truncation of the infinite sum over $\kappa$ in Eq.
(\ref{expan}) has been made to include at least 30 partial waves
for both initial and final electron, as well as for the outgoing
proton in DWIA. The radial integrals in Eq. (\ref{eqb8}) have been
carried out numerically up to typically 15 fm in the nuclear
coordinate and 30 fm in the electron coordinate. This gives a
good compromise in optimizing both numerical accuracy and computing
time. The numerical accuracy has been checked by comparison to PWBA
of the results obtained in DWBA for $Z\rightarrow 0$.
The estimated numerical error amounts to less than 2\% in the
spectroscopic factors \cite{Udi93}.

\section{Results}
We present our results in terms of reduced cross-sections
$\rho (P_m)$ for selected $E_m$ values (i.e., for selected
single-particle shells), defined by
\begin{equation}
\rho(P_m)=\int_{\Delta E_m} dE_m\ \left[ {\sigma_{ep}|\vec{P_F}|E_F}
\right] ^{-1} \frac{d\sigma}{d E_F d\epsilon_f
d\Omega_F d\Omega_f} \; ,
\label{reduc}
\end{equation}
as  functions of the missing momentum
($\vec{P}_m=\vec{P}_A-\vec{P}_{A-1}$). Experimentally, the integral
over the missing energy ($E_m=M_{A-1}+M-M_A$) is taken over the
interval $\Delta E_m$ that contains the peak of the transition under
study. In our calculations of $\rho(P_m)$, we take for $\sigma_{ep}$
in Eq. (\ref{reduc}) the same expression used by the
experimentalists, i.e., we use the expression
$\sigma_{cc1}$ given by Eq. (17) of Ref. \cite{deF83}.

We present results for the $3s_{1/2}$ and $2d_{3/2}$ shells in
$^{208}$Pb and for the $2s_{1/2}$ and $1d_{3/2}$ shells in $^{40}$Ca
and compare with data in parallel kinematics
($\vec{q} \parallel \vec{P}_F$). We have chosen to study these
shells for several reasons: {\it i)} they correspond to
experimentally well separated peaks, {\it ii)} for these doubly
magic nuclei the theoretical description is simpler and there are
available optical potentials, {\it iii)} this choice allows us to
study different mass regions where the Coulomb distortion of the
electron waves is expected to play a different role, and {\it iv)}
for these shells we can compare our results with experimental data
as well as with other relativistic and/or nonrelativistic theoretical
results. Unless otherwise specified the results presented correspond
to parallel kinematics ($\vec{q} \parallel \vec{P}_F$). All the
calculations have been done for a fixed value of the kinetic energy
of the outgoing proton ($T_F=100$ MeV). We also take a fixed value
of the incoming electron energy, $\epsilon_i=412$ MeV (375 MeV) for
$^{208}$Pb ($^{40}$Ca).

\subsection{$3s_{1/2}$ shell of  $^{208}$Pb}

We discuss results on the $3s_{1/2}$ shell in detail because this
shell is the most extensively studied in ($e,e'p$), both
experimentally and theoretically.

To start with, it is worth clarifying the situation with regards to
the disagreement between the PWBA results in Refs. \cite{McD90} and
\cite{Jin92}.  As pointed out in Ref. \cite{Jin92}, both calculations
being the same in PWBA should lead to the same results. We found
that the reason for the disagreement is due to the fact that
different wave functions were used for the bound proton. In our
calculations we used the code by Horowitz and Serot (TIMORA) with
the standard set of initial conditions and obtained a result (see
dashed line in Fig. 2a) in agreement with that in Ref. \cite{Jin92}.
With the same code but changing the initial conditions to those
used by McDermott \cite{McD90,Joeprivate}, we find a result (see
dotted line in Fig. 2a) that agrees with the result in Ref.
\cite{McD90}.  As indicated in the figure, the two above mentioned
results have been obtained with the operator $cc2$, which is the
operator always used in Refs. \cite{McD90,Jin92}.

In what follows we use always for the bound proton the wave function
obtained with the standard set of initial conditions. This wave
function corresponds to a binding energy of 5.7 MeV.

In Fig. 2a we also show our result (solid line) in PWBA corresponding
to the choice $cc1$ for the current operator. As seen in this figure
the PWBA results obtained with $cc1$ (solid) and $cc2$ (dashed)
current operators are very close. The main difference seen at the
peaks is less than 8$\%$, the $cc1$ result being larger.

In Fig. 2b we show our DWBA results obtained with the $cc1$ operator
(solid line) together with the DWBA results of Ref. \cite{Jin92}
(dashed line) that was obtained with the $cc2$ operator. Also shown
in this figure is the result obtained with the code of McDermott,
used in Ref. \cite{McD90}, that makes use of the HCA (dotted line).
In this figure, the three curves are obtained using the same
relativistic optical potential \cite{Hama} and also the same wave
function for the bound proton. One can see that our results (solid
line) and those from Ref. \cite{Jin92} (dashed line) are similar and
show up the same qualitative changes with respect to the
corresponding PWBA results shown in Fig. 2a. These changes are:
{\it i)} a shift in $P_m$, {\it ii)} an increase at the maxima, and
{\it iii)} a filling of the minimum at $P_m\sim 150$ MeV. Note that
the result obtained  within HCA (dotted line in Fig. 2b) is not
able to reproduce the two last mentioned effects (focussing effects)
but it reproduces adequately the shift in $P_m$.

In Fig. 3a we show in more detail the comparison between our PWBA
and DWBA results shown by the dotted and solid lines, respectively.
Also shown in Fig. 3a is the PWBA result with an effective $q$-value
(dashed line), $q_{eff}=|\vec{k}_{i,eff}-\vec{k}_{f,eff}|$, where
the electron initial and final effective momenta are given by
\begin{equation}
\vec{k}_{eff} = \vec{k} + f_{c}\frac{Z\alpha}{R}\frac{\vec{k}}
{|\vec{k}|}
\label{shift}
\end{equation}
with $R=1.1\ A^{1/3}$ fm and $f_c=1.35$. This $f_c$ value has been
adjusted  to get the same Coulomb potential at the origin that
corresponds to the charge distribution of the nuclear target
obtained from the relativistic calculation, which agrees with the
experimental one \cite{deV}. This value is somewhat smaller than
the value $f_c=3/2$, corresponding to a spherical uniform charge
distribution. In parallel kinematics, since $P_m=q-P_F$ is varied
by varying $q$, a displacement in $q$ produces a displacement in
$P_m$. As seen in Fig. 3a, with this $f_c$ value the shift in
$P_m$ caused by electron Coulomb distortion is well accounted for.

It should be pointed out that the effect of Coulomb distortion
shows up differently in perpendicular kinematics (constant $q$
and $\omega$). To illustrate this point we show in Fig. 3b the
comparison between our results in PWBA and DWBA in perpendicular
kinematics. In this case there is no observable effect of
displacement with $P_m$ but there is a reduction of the
cross-section in DWBA at $P_m\sim 0$ that, as seen in the figure,
is accounted for replacing $q$ by $q_{eff}$ in PWBA. The reduction
is due to the fact that the maximum of the bound nucleon wave
function  ($s$-wave) at $P_m=0$ is not reached when $q$ is replaced
by $q_{eff}$. The actual effect of the focussing is of the same
order in perpendicular kinematics than in parallel kinematics.
On the other hand changing $q$ into $q_{eff}$ produces quite a
different effect in perpendicular kinematics because in this case
$P_m$ is varied keeping $q$ constant and varying the angle between
$\vec{q}$  and $\vec{P}_F$. In what follows only the case of
parallel kinematics is considered.

The focussing effects, causing the filling of the minimum and
increase of the maxima in Fig. 3a, are summarized in a quantitative
way in Table I. In this table we quote the ratios between the
reduced cross-sections calculated in DWBA and in PWBA at the two
maxima. We compare the ratios obtained in this work with the ratios
deduced from other relativistic and nonrelativistic calculations
previously reported. The nonrelativistic results in Table I
correspond to the code DWEEPY \cite{GP}. This code treats the
Coulomb distortion in an approximate way making use of an expansion
in powers of $(Z\alpha /k)$ to first order (DWEEPY1) or to second
order (DWEEPY2). DWEEPY2 is the most commonly used approximation in
the nonrelativistic analyses of the data. One can see in Table I
that our results and those of Ref. \cite{Jin92}, which also treat
Coulomb distortion in an exact way, are in very good agreement. The
results obtained with the HCA code of McDermott
\cite{McD90,Joeprivate}  do not reproduce adequately the focussing
effect at the peaks. Among the nonrelativistic calculations,
DWEEPY1 overestimates largely the effect of focussing and DWEEPY2
gives still an overestimation. The spectroscopic factors are very
sensitive to focussing effects because these factors are derived by
scaling the theoretical reduced cross-section to the experimental
one. As seen in Table I the various approaches to treat the electron
Coulomb distortion fail to account for the focussing effect obtained
with the exact DWBA calculation, which is non-negligible in
$^{208}$Pb. Therefore, DWBA calculations are necessary to deduce
reliable spectroscopic factors in heavy nuclei such as lead.

In comparing relativistic and nonrelativistic calculations, it is
also important to know how  different approximations for the proton
wave functions affect the reduced cross-section. To this end we
compare in Fig. 4 the results of relativistic and nonrelativistic
calculations in PWBA because in this case there is no effect from
electron Coulomb distortion and the differences come only from the
various approximations at the nuclear vertex.

In Fig. 4 the dotted and dashed lines show the PWIA results of the
nonrelativistic and relativistic calculations, respectively. The
solid and short-dashed lines show the results in DWIA of
relativistic and nonrelativistic calculations. For the relativistic
calculations we use the nuclear current operator $J_{cc2}$ in Eq.
(\ref{cc2}) and the standard TIMORA solution \cite{HSbook} for the
relativistic bound proton wave function. For the nonrelativistic
bound proton wave function we use the upper component of the
relativistic one, properly normalized, in the standard
nonrelativistic form of the nucleon current, based on $cc2$
\cite{McV}. The nonrelativistic optical potential has been taken
from set II in Table 2.1 of Ref. \cite{Quint}, which was determined
following the procedure of Ref. \cite{Blok87}. This set gives a
good fit to elastic proton scattering data \cite{Sch82} from
$^{208}$Pb at an energy of 98 MeV (which roughly corresponds to the
proton energy involved in the $(e,e'p)$ experiments \cite{Quint}).
On the other hand, the relativistic optical potential used here was
determined \cite{Hama} from a global fit to elastic proton
scattering data from spherical nuclei with mass numbers
$40\le $ A $\le $208 in a wide range of projectile energies from
65 MeV to 1040 MeV.

As seen in Fig. 4, both relativistic and nonrelativistic calculations
give practically the same result in PWIA. The same is true when a
Woods-Saxon potential is used for the bound nucleon \cite{Udi93}.
This means that the use of nonrelativistic bound nucleon wave
functions does not produce a significant change in the $P_m$ range
considered.

The differences between relativistic and nonrelativistic calculations
are much more noticeable in DWIA. As seen in Fig. 4, the shape is
similar in both DWIA calculations but the relativistic potential
leads to stronger absorption than the nonrelativistic one
(about a 20$\%$ difference at the peaks). This is at first sight
surprising since both potentials reproduce well the elastic
scattering of protons from this nucleus at proton energies of about
100 MeV. Therefore, the difference seen between the DWIA results in
Fig. 4 must be attributed to details of the optical potential to
which the elastic proton scattering at this energy is not sensitive.

As it is well known \cite{HS} the relativistic S-V potential gives
rise to non-local terms in the nonrelativistic reduction. The effect
of these terms in the ($e,e'p$) reaction was investigated by Boffi
and collaborators \cite{Boffi}. The authors of Ref. \cite{Boffi}
concluded that these terms only affect the inelastic processes, where
the nuclear interior is important, producing about a 15$\%$ increase
of the absorption due to final state interactions in the ($e,e'p$)
reaction. This conclusion is in agreement with the increased
absorption observed in Fig. 4 in going from the nonrelativistic DWIA
to the relativistic DWIA. It would be interesting to see to what
extent the nonrelativistic reduction of the S-V optical potential
\cite{Hama} may lead to a similar absorption as the relativistic
DWIA. Work along these lines is in progress.

\subsection{Comparison with experiment}

The main results of this work are presented in Table II and in
Fig. 5, where we compare our relativistic DWBA results to the
experimental data, in parallel kinematics. Our theoretical results
are given by the solid lines in Fig. 5 and have been scaled by the
spectroscopic factors presented in Table II.

Our spectroscopic factors for the $3s_{1/2}$ and $2d_{3/2}$ shells
in $^{208}$Pb, and for the $2s_{1/2}$ and $1d_{3/2}$ shells in
$^{40}$Ca, are given in the third and fourth columns of Table II,
corresponding to calculations with $cc1$ and $cc2$ operators,
respectively. They have been obtained by scaling our theoretical
results on $\rho (P_m)$ to the experimental data from Refs.
\cite{Quint,Kra89} shown in Fig. 5. For each shell the overall
scale factor (spectroscopic factor) has been obtained by means of
an error weighted least squares procedure. The quoted errors in
our spectroscopic factors include both statistical and systematical
errors in experimental data \cite{Quint,Kra89}. Also given for
comparison in Table II are the spectroscopic factors obtained from
other relativistic and nonrelativistic calculations. The
spectroscopic factors reported by Jin {\em et al.} \cite{Jin92}
have been obtained with a fully relativistic DWBA formalism similar
to the one used in this work, using the $cc2$ current operator. As
discussed in the previous section, in the case of the $3s_{1/2}$
shell in $^{208}$Pb the authors of Ref. \cite{Jin92} use the same
bound nucleon wave function used here. For $^{40}$Ca the bound
nucleon wave functions used in Ref. \cite{Jin92} are different
\cite{Jin91}  from the standard TIMORA solutions used here. In
column six of Table II we quote the spectroscopic factors obtained
by McDermott \cite{McD90} using also the $cc2$ operator. The last
column of Table II contains  the spectroscopic factors obtained
with the nonrelativistic analyses reported in Refs.
\cite{Quint,Kra89}.

As seen in Table II, in general, the spectroscopic factors obtained
with the $cc2$ current operator  are somewhat larger than those
obtained with $cc1$. For $^{208}$Pb our $cc2$ results differ by
less than 10\%  from the $cc1$ results and are in agreement with
the corresponding result by Jin {\em et al.} \cite{Jin92}. As
discussed in the previous section, the result by McDermott
\cite{McD90} does not contain the correct focussing and corresponds
to a different $3s_{1/2}$ relativistic wave function leading to a
smaller spectroscopic factor. Nevertheless, it is important to
remark that all the results of the relativistic calculations for
$^{208}$Pb are in agreement within a 10\% and are compatible with
theoretical predictions \cite{Mah91,Ma91,Pan84}, while the previous
results from the nonrelativistic analyses are clearly too low.
These too small spectroscopic factors in the nonrelativistic
analyses of Refs. \cite{Quint,Kra89} result from too much focussing
of the electron waves and too little absorption of the outgoing
proton wave, compared to the relativistic analyses, as was discussed
in detail in the previous section (see Table I and Fig. 4).

It should be stressed here that in our analysis the determination
of the spectroscopic factors has no free parameters. All of the
parameters entering in the relativistic potentials were obtained
from independent considerations \cite{Hama,HS}. Taking this into
account it is remarkable the good quality of the fits to the
experimental data seen in Fig. 5a for the two shells in $^{208}$Pb.
To have a measure for the quality of the fit, we consider the
parameter $Q$ defined as the $\chi ^2$-value divided by the degrees
of freedom (number of data involved minus one) in the  determination
of the spectroscopic factor. In our cases $Q < 3$ corresponds
to a good quality of the fit while $Q > 5$ corresponds to a
poor fit. The $Q$-values obtained for $^{208}$Pb are $Q < 3$
for both shells ($3s_{1/2}, 2d_{3/2}$) and for both types of current
operators. Our results in Fig. 5a correspond to the calculations
with the $cc1$ operator, using the spectroscopic factors in column
three of Table II. Similar results are obtained using the $cc2$
operator and the spectroscopic factors of column four in Table II.

In the case of $^{40}$Ca the focussing effect  is negligible and
the only noticeable effect of electron Coulomb distortion is a shift
in $P_m$. This is illustrated in Fig. 5b for the $2s_{1/2}$ orbital
where we plot the DWBA result (solid line) and the result in PWBA
with $q_{eff}$ (dotted line), both using the $cc1$ operator and the
same spectroscopic factor ($S=0.44$, see Table II).  The effective
momentum corresponds to $f_c=1.35$ in Eq. (\ref{shift}).  As seen
in the figure there is no significant difference between the
results in DWBA and in PWBA with $q_{eff}$. Although not shown in
the figure, the same is true with regards to DWBA and PWBA
calculations for the $1d_{3/2}$ orbital in $^{40}$Ca.

As seen in Fig. 5b the agreement between our theoretical results
for the  $2s_{1/2}$ and for the $1d_{3/2}$ shells in $^{40}$Ca
with the $cc1$ current operator is not as good as that obtained for
the shells in $^{208}$Pb. If we use instead the $cc2$ operator, the
agreement is not improved for the $2s_{1/2}$ shell, but it improves
considerably for the $1d_{3/2}$ shell. This is seen in Fig. 5b where
we show our results for the $1d_{3/2}$ shell with the $cc1$ and the
$cc2$ operators by solid and dashed lines respectively. For the
case of the $2s_{1/2}$ shell the $Q$ value is larger than 7 in all
cases, showing the poor quality of the fit. For the $1d_{3/2}$ shell
the $Q$ value is almost 5 in the case of the $cc1$ operator and is
less than 2 in the case of the $cc2$ operator.

It is important to realize that all the cases where our spectroscopic
factors in columns three and four of Table II are low correspond to
situations in which the shape of the theoretical curve does not match
well the trend in the experimental data and the $Q$ values are large,
i.e., the fits are poor. This is particularly the case for the
$2s_{1/2}$ shell in $^{40}$Ca, where our spectroscopic factor is as
low as that obtained in the nonrelativistic analysis of Ref.
\cite{Kra89}. For the shell $1d_{3/2}$ in $^{40}$Ca, where the fit
with the $cc2$ operator is very good, we get a spectroscopic factor
$S=0.76$, similar to that found in $^{208}$Pb, and larger than the
one obtained in the nonrelativistic analysis \cite{Kra89}.

In the case of $^{40}$Ca the comparison of our results with those
of Ref. \cite{Jin92} is not as meaningful as in the case of
$^{208}$Pb for the two following reasons: {\em i)} the bound nucleon
wave functions used in Ref. \cite{Jin92} were different \cite{Jin91}
than the standard TIMORA solutions used here, and {\em ii)} the
spectroscopic factors in Ref. \cite{Jin92} were derived by visual
fitting. However it is interesting to see that the result of Jin
{\em et al.} \cite{Jin92} agrees with the one we obtain for the
$1d_{3/2}$ shell with the $cc2$ operator, where we consider our
deduced spectroscopic factor to be reliable.

\section{Concluding remarks}

We have analyzed the quasielastic $(e,e'p)$ reaction within a fully
relativistic formalism treating the Coulomb distortion of the
electrons in an exact way. This analysis is expected to yield more
reliable spectroscopic factors than previous analyses where the
electron Coulomb distortion was not fully taken into account
\cite{GP,McD90}.

We present results for reduced cross-sections as functions of the
missing momentum ($P_m$), corresponding to proton knock-out from
the outermost shells in $^{40}$Ca and in $^{208}$Pb. We study the
effects purely due to electron Coulomb distortion as well as the
effects due to the relativistic treatment of the initial (bound)
and final (distorted) nucleon wave functions. We also study the
effects of using different nucleon current operators.

In parallel kinematics, Coulomb distortion produces two effects:
{\em i)} displacement of the cross-section towards higher $P_m$
values that can be simulated by the use of an effective momentum
transfer $q_{eff}$ in PWBA and {\em ii)} focussing that shows up
mainly in the maxima and minima of the reduced cross-section.

The focussing effect plays an important role in the determination
of the spectroscopic factors in $^{208}$Pb.  In $^{40}$Ca the
focussing effect  is much smaller than in $^{208}$Pb and has no
influence in the extraction of spectroscopic factors, but the
displacement effect is still sizeable. Exact DWBA calculations
are worthwhile even in medium nuclei like $^{40}$Ca in order to
determine the precise magnitude of this displacement and to avoid
introducing additional parameters.

In perpendicular kinematics, we do not find displacement in $P_m$
but there is an apparent increase of focussing (or antifocussing)
at $P_m\sim 0$, caused by the effective momentum transfer.

We find that the main difference between the results obtained with
our relativistic treatment of the nuclear vertex and with the usual
nonrelativistic treatment \cite{Quint,Kra89} is due to the fact
that the relativistic optical potential \cite{Hama} produces more
absorption in the ($e,e'p$) reaction. This gives rise to higher
spectroscopic factors in much better agreement with theoretical
predictions \cite{Mah91,Ma91,Pan84}. In our view this is a strong
point in favour of the relativistic S-V models that are able to
account for both elastic proton-nucleus and ($e,e'p$) data
simultaneously and confirms the fact that the relativistic
phenomenology is in general superior to the usual nonrelativistic
phenomenology in covering a large variety of experimental
information on proton-nucleus scattering and particularly in what
concerns spin rotation functions (see Ref. \cite{Fes92} and
references therein).

We also find that the off-shell electron-proton cross-section is
less sensitive to the choice of the nucleon current operators when
using free Dirac spinors \cite{deF83,Cab93} than when using bound
Dirac spinors. Thus, one should be cautious when estimating the
uncertainty associated to the current operator from results based
only on free spinors. Actually, further work is needed to clarify
the situation with regards to the best choice of the nucleon
current operator in the fully relativistic treatment of the
($e,e'p$) reaction.

The spectroscopic factors have been obtained by scaling to
experimental data the calculated reduced cross-sections without
introducing any free parameter. For the $3s_{1/2}$ and $2d_{3/2}$
shells in $^{208}$Pb we find spectroscopic factors between 0.65
and 0.73 depending on the shell and on the nucleon current operator.
High quality fits to the experimental reduced cross-sections are
obtained in all the cases. These spectroscopic factors are in
agreement with theoretical predictions and with the results obtained
for the $3s_{1/2}$ shell in $^{208}$Pb by Jin {\em et al.}
\cite{Jin92} from a fully relativistic analysis analogous to the one
carried out here. They are considerably larger than the
spectroscopic factors obtained from the nonrelativistic analyses
\cite{Quint}.

For $^{40}$Ca we only obtain a high quality fit to the data for the
$1d_{3/2}$ shell when using the $cc2$ operator. In this case the
value of the extracted spectroscopic factor is 0.76, similar to
those obtained in $^{208}$Pb. For the $2s_{1/2}$ shell the quality
of the fit is very poor with both $cc1$ and $cc2$ operators and,
therefore, we cannot consider the deduced spectroscopic factors as
being reliable. In our view, if the shape of the cross-section is
not well reproduced, the derived spectroscopic factor is not
meaningful. For this particular orbital the wave function obtained
from the TIMORA code used here does not reproduce the experimentally
observed shape of the reduced cross-section. In comparing with the
results by Jin {\em et al.} \cite{Jin92} for these two orbitals
in $^{40}$Ca, one should keep in mind that there are two main
differences between their calculations and ours; {\em i)} the
results reported in Ref. \cite{Jin92} were obtained using a
phenomenological potential fitted to reproduce single-particle
properties in $^{40}$Ca \cite{Jin91}; {\em ii)} their spectroscopic
factors are obtained by visual fit and no quality of the fit is
reported to compare with ours. Nevertheless, in the case of the
shell $1d_{3/2}$ with the $cc2$ operator, where the quality of our
fit is good, our spectroscopic factor agrees with that of Ref.
\cite{Jin92}.

In this work we made no attempt to fit to experiment the r.m.s.
radii and binding energies of the orbitals. Thus, the fact that we
get high quality fits for all the orbitals studied, except the
$2s_{1/2}$ in $^{40}$Ca, can be considered as a success of the
relativistic analysis. It would be interesting to explore whether
more elaborated relativistic S-V models, including non linear terms,
may produce results in better agreement with experimental data on
reduced cross-sections for the $2s_{1/2}$ shell in $^{40}$Ca.

\acknowledgements

We are grateful to J. McDermott for providing us with his HCA code
and for useful discussions and to L. Lapik\'as and G. van der
Steenhoven for communication of experimental data collected at
NIKHEF-K. One of us (J.M.U.) also thanks Y. Jin for helpful comments.
This work has been partially supported by DGICYT (Spain) under
contract 92/0021-C02-01.

\begin{figure}
\caption{Schematic picture of the $(e,e'p)$ process.}
\end{figure}
\begin{figure}
\caption{Reduced cross-sections for the shell $3s_{1/2}$ of
$^{208}$Pb in DWIA. (a) Comparison of PWBA results with $cc1$ and
$cc2$ operators. The dashed and dotted lines correspond to different
choices of the $3s_{1/2}$ wave function. (b) Comparison of DWBA
results with $cc1$ and $cc2$ operators. See text.}
\end{figure}
\begin{figure}
\caption{Reduced cross-sections for the shell $3s_{1/2}$ of
$^{208}$Pb in DWIA corresponding to parallel (a) and perpendicular
(b) kinematics. The results obtained in DWBA are compared to the
results in PWBA with and without effective momentum transfer.}
\end{figure}
\begin{figure}
\caption{Comparison of PWBA results from relativistic (r.) and
nonrelativistic (n.r.) calculations in DWIA and PWIA for the
$3s_{1/2}$ shell in $^{208}$Pb.}
\end{figure}
\begin{figure}
\caption{Theoretical and experimental reduced cross-sections for the
$3s_{1/2}$ and $2d_{3/2}$ shells in $^{208}$Pb (a) and for the
$2s_{1/2}$ and $1d_{3/2}$ shells in $^{40}$Ca (b). Experimental data
are from Refs. \protect\cite{Quint,Kra89}. Solid lines are the
results of DWBA+DWIA calculations with $cc1$ operator, scaled by the
spectroscopic factors in the third column of Table II. The dotted
line in (b) corresponds to the result with PWBA+DWIA with the $cc1$
operator. The dashed line in (b) corresponds to the result with the
$cc2$ operator.}
\end{figure}
\newpage
\mediumtext

\begin{table}
\caption{Ratio between DWBA and PWBA reduced cross-sections
at the two maxima in the  $3s_{1/2}$ shell of $^{208}$Pb
( $T_F=100$ MeV and $\epsilon_i=412$ MeV with parallel kinematics.)
\label{foc} }
\begin{tabular}{lccccc}
              & This work & Jin {\em et al.} \cite{Jin92} &  HCA
 & DWEEPY1 \tablenotemark[1] & DWEEPY2\tablenotemark[1] \\ \tableline
First Maximum & 1.08  & 1.08      & 0.98  & 1.92 & 1.21  \\
Second Maximum & 1.14 & 1.13      & 1.03  &  $-$ & $-$    \\
\end{tabular}
\tablenotetext[1]{Extracted from Ref. \protect\cite{GP}}
\end{table}

\begin{table}
\caption{Comparison of our spectroscopic factors obtained with the
$cc1$ and $cc2$ operators and with
other  relativistic and nonrelativistic analyses.
\label{factors} }
\begin{tabular}{llccccc}
           &            & cc1      & cc2     & Jin {\em et al.}
\cite{Jin92}& McDermott\cite{McD90}& nonrel. \\ \tableline
$^{208}$Pb & $3s_{1/2}$ &  0.65(4) & 0.70(4) & 0.71     & 0.65     &
0.50(5) \tablenotemark[1] \\
           & $2d_{3/2}$ &  0.66(4) & 0.73(4) & $-$      & $-$      &
0.53(4) \tablenotemark[1] \\ \tableline
$^{40}$Ca  & $2s_{1/2}$ &  0.44(3) & 0.51(3) & 0.75     & $-$      &
0.50(6) \tablenotemark[2] \\
           & $1d_{3/2}$ &  0.60(3) & 0.76(4) & 0.80     & 0.66     &
0.65(5) \tablenotemark[2] \\
\end{tabular}
\tablenotetext[1]{From Ref. \protect\cite{Quint}}
\tablenotetext[2]{From Ref. \protect\cite{Kra89}}

\end{table}


\begin{references}
\bibitem{Mah91}
C. Mahaux and R. Sartor, Adv. in Nucl. Phys. {\bf 20}, 1 (1991).
\bibitem{Ma91}
Z.Y. Ma and J. Wambach, Phys. Lett. B{\bf 256}, 1 (1991).
\bibitem{Pan84}
V.R. Pandharipande, C.N. Papanicolas and J. Wambach,  Phys. Rev.
Lett. {\bf 53}, 1133 (1984).
\bibitem{sumrule}
J.M. Cavedon {\em et al.}, Phys. Rev. Lett. {\bf 49}, 978 (1982);
E.N.M. Quint {\em et al.}, Phys. Rev. Lett. {\bf 57}, 186 (1986);
E.N.M. Quint {\em et al.}, Phys. Rev. Lett. {\bf 58}, 1088 (1987).
\bibitem{Sic91}
I. Sick and P.K.A. de Witt Huberts, Comments Nucl. Part. Phys.
{\bf 20}, 177 (1991).
\bibitem{Quint}
E.N.M.  Quint, Ph. D. Thesis, University of Amsterdam (1988).
\bibitem{Kra89}
G.J. Kramer,  Phys. Lett. B{\bf 227} 199 (1989);
Ph.D. Thesis, University of Amsterdam (1990).
\bibitem{deW90}
P.K.A. de Witt Huberts, J. Phys. G {\bf 16}, 507 (1990).
\bibitem{Fru85}
S. Frullani and J. Mougey, Adv. in Nucl. Phys. {\bf 14}, 1 (1985).
\bibitem{GP}
C. Giusti and F. Pacati, Nucl. Phys. {\bf A473}, 717 (1987);
Nucl. Phys. {\bf A485}, 461 (1988);
M. Traini, Phys. Lett. B{\bf 213},1 (1988).
\bibitem{McD90}
J.P. McDermott,  Phys. Rev. Lett. {\bf 65}, 1991 (1990).
\bibitem{Jin92}
Y. Jin, D.S. Onley and L.E. Wright,  Phys. Rev. C{\bf 45}, 1311
(1992).
\bibitem{Udi93}
J.M. Ud\'{\i}as,  Ph.D. Thesis, Universidad Aut\'{o}noma de Madrid
(1993).
\bibitem{BD64}
J.D. Bj\"orken and S.D. Drell, {\it Relativistic Quantum
Mechanics} (Mc Graw-Hill, N.Y. 1964).
\bibitem{UbeRose}
H. Uberall, {\it Electron Scattering from Complex Nuclei} (Academic
Press, N.Y. 1971);
M.E. Rose,  {\it Relativistic Electron Theory} (Wiley, 1961).
\bibitem{Yen}
D.R. Yennie, R.N. Wilson and D.G. Ravenhall, Phys. Rev. {\bf 95},
500 (1954).
\bibitem{deV}
H. de Vries, C.W. de Jager and C. de Vries, At. Data Nucl. Data
Tables {\bf 36}, 495 (1987).
\bibitem{HS}
C.J. Horowitz and B.D. Serot, Nucl. Phys. {\bf A368}, 503 (1981);
Phys. Lett. B{\bf 86},  146 (1979).
\bibitem{HSbook}
C.J. Horowitz, D.P. Murdock and B.D. Serot,
{\it Computational Nuclear Physics}, (Springer-Verlag, Berlin, 1991).
\bibitem{Hama}
S. Hama, B.C. Clark, E.D. Cooper, H.S. Sherif and R.L. Mercer,
Phys. Rev. C{\bf 41}, 2737 (1990);
S. Hama, E.D. Cooper, B.C. Clark and R.L. Mercer,
in {\em Global Dirac Phenomenology for Proton--Nucleus Elastic
Scattering}, Ohio State University, {\em preprint} August 1992.
\bibitem{deF83}
T. de Forest, Jr.,  Nucl. Phys. {\bf A392}, 232 (1983).
\bibitem{Cab93}
J.A. Caballero, T.W. Donnelly and G.I. Poulis, Nucl. Phys. {\bf A555},
709 (1993).
\bibitem{Joeprivate}
J.P. McDermott, {\em private communication.}
\bibitem{McV}
K.V. McVoy and L. van Hove, Phys. Rev. {\bf 125}, 1034 (1962).
\bibitem{Blok87}
H.P. Blok, L.R. Kouw, J.W.A. den Herder, L. Lapik\'as and
P.K.A. de Witt Huberts, Phys. Lett. B{\bf 198}, 4 (1987).
\bibitem{Sch82}
P. Schwandt, H.O. Meyer, W.W. Jacobs, A.D. Bacher, S.E. Vigdor, M.D.
Kaitchuck and T.R. Donoghue, Phys. Rev. C{\bf 26}, 55 (1982).
\bibitem{Boffi}
S. Boffi, C. Giusti, F.D. Pacati and F. Cannata, Nuovo Cimento
{\bf 98}, 291 (1987).
\bibitem{Jin91}
Y. Jin, Ph.D. Thesis, Ohio University (1991).
\bibitem{Fes92}
B.C. Clark, R.L. Mercer and P. Schwandt, Phys. Lett. B{\bf 122}, 211
(1983); H. Feshbach, {\it Theoretical Nuclear Physics} (John Wiley
\& Sons, 1992).
\end{references}
\end{document}